\documentclass{SciPost}

\hypersetup{
    colorlinks,
    linkcolor={red!50!black},
    citecolor={blue!50!black},
    urlcolor={blue!80!black}
}

\usepackage[bitstream-charter]{mathdesign}
\DeclareSymbolFont{usualmathcal}{OMS}{cmsy}{m}{n}
\DeclareSymbolFontAlphabet{\mathcal}{usualmathcal}

\fancypagestyle{SPstyle}{
\fancyhf{}
\lhead{\colorbox{scipostblue}{\bf \color{white} ~SciPost Physics Proceedings }}
\rhead{{\bf \color{scipostdeepblue} ~Submission }}

\fancyfoot[C]{\textbf{\thepage}}
}

\usepackage{caption}
\usepackage{subcaption}
\usepackage{amsmath}

\newcommand{\MADGRAPH} {\textsc{MadGraph}}
\newcommand{\MCATNLO} {\textsc{mc@nlo}}

\begin{document}

\pagestyle{SPstyle}

\begin{center}{\Large \textbf{\color{scipostdeepblue}{
First simultaneous measurement of single and pair production of top quarks in association with a Z boson at the LHC
}}}\end{center}

\begin{center}\textbf{
Konstantin Sharko\textsuperscript{1$\star$} on behalf of the CMS collaboration
}\end{center}

\begin{center}
{\bf 1} Deutsches Elektronen-Synchrotron DESY
\\[\baselineskip]
$\star$ \href{mailto:konstantin.sharko@cern.ch}{\small konstantin.sharko@cern.ch}
\end{center}

\section*{\color{scipostdeepblue}{Abstract}}
\textbf{\boldmath{
The first simultaneous measurement of single and pair production of top quarks in association with a Z boson (tZq, $t$WZ and $t\bar{t}$Z) is presented, including both inclusive and differential cross sections.
A multiclass neural network is used to separate the signal and the backgrounds. 
Compared to previous studies, the simultaneous measurement is less dependent on the signal modeling assumptions and improves the sensitivity to new physics scenarios, as it enables to constrain possible deviations from the standard model across different processes.
}}

\vspace{\baselineskip}

\section{Introduction}
\label{sec:intro}

Precise experimental measurements provide tests to the Standard Model (SM) predictions and identify possible deviations that might lead to the discovery of new physics.
In particular, the measurement of top quark production in association with a Z boson allows to directly probe couplings between these two particles.
Processes involving top quarks and Z bosons are significant backgrounds to other SM processes such as Higgs boson production in association with a top quark. 
In the SM couplings between quarks and Z bosons are predicted to be quark-flavor conserving.
Possible physics beyond the SM could induce measurable deviations from this behavior.

At the TOP Workshop 2024 the first simultaneous measurement of the single and pair production of top quarks in association with a Z boson was presented.
The preprint version of the paper is available at~\cite{colombina_2024}.

\section{Event selection and classification}
\label{sec:selection}
The analysis is done with the dataset of proton-proton collisions of the CMS experiment~\cite{CMS_2008, CMS_2024} taken at Run-2 of Large Hadron Collider (LHC) at a center-of-mass energy of 13~TeV. 
Correspondent integrated luminosity is equal to 138 fb$^{-1}$.

For all events exactly three leptons are required. 
Two of them must have the same flavor, opposite charge and an invariant mass between 70 and 110~GeV to pick Z decay events. 
The third one is taken to pick the W boson decay, originating from t quark.
The pseudorapidity selection criteria are $\left|\eta\right| <$~2.5 for electrons and $\left|\eta\right| <$~2.4 for muons.
The criteria for transverse momenta of the three leptons is $p_T >$~25, 15 and 10~GeV.
Jets kinematic requirements are $p_T >$~25 GeV and $\left|\eta\right| <$~5.0.
Spatial separation between jets and each lepton is $\Delta R = \sqrt{\delta \eta^2 + \delta \phi^2} >$~0.4. 
At least two jets are required and at least one of them has to be tagged as b quark.

The main contribution to the nonprompt background comes from $t\bar{t}$ and Drell–Yan processes
where one lepton candidate is misidentified. 
To estimate its impact a method similar to that from~\cite{tZ_2022} is used.
A lepton misidentification rate is determined separately for electrons and muons in a region enriched with backgrounds.
This information is used to determine a transfer factor that is applied to events in the data with three leptons, of which at least one does not fulfill a multivariate classifier criteria. 
The obtained distribution describes the expectation for the nonprompt background in the signal region.
To validate the estimation of the nonprompt lepton contribution, events outside the Z boson resonance
region are selected, $\left| m(ll) - m(Z)\right|$~<~20 GeV. 
In this region, the contribution from events with ''nonprompt'' leptons is enhanced. 

Based on the selection a deep neural network (DNN) is used to split the events into 3 categories: $t\bar{t}$Z~+~$t$WZ, $t$Zq and background.
The DNN input consist of 26 variables, including jet, lepton and reconstructed top kinematic observables, final state charge and jet multiplicity.
All the events are split into two equal sized subsamples.
The first subsample is then further split: 80\% of it is used for training of the model and 20\% for it's testing.
The second subsample is used to evaluate the model and to build the output distributions.
The output score is normalized such that for each event their sum gives unity.
For use in a signal extraction fit each event is then assigned to a category, in which it has the highest output score.

\section{Results}

The cross sections are measured using a profile likelihood-ratio scan, in which for each cross section hypothesis the optimal nuisance parameters are determined from a fit. 
For the simultaneous inclusive measurement two parameters of interest are used in the simultaneous measurement: $t\bar{t}$Z~+~$t$WZ and $t$Zq cross sections.
For the simultaneous measurement of the differential cross sections, two corresponding parameters are used in each bin of the differential measurement.

For the inclusive cross sections, besides the event classifier maximum scores distributions, the event selection includes a category of events with four leptons, and a category of events without
b jets. 
The region with four leptons and at least one b jet has a high purity in $t\bar{t}$Z events. 
The region without b jets is enriched in WZ background events and is useful as a control region. 

\begin{figure}[!hbtp]
	\centering
	\includegraphics[width=0.5\textwidth]{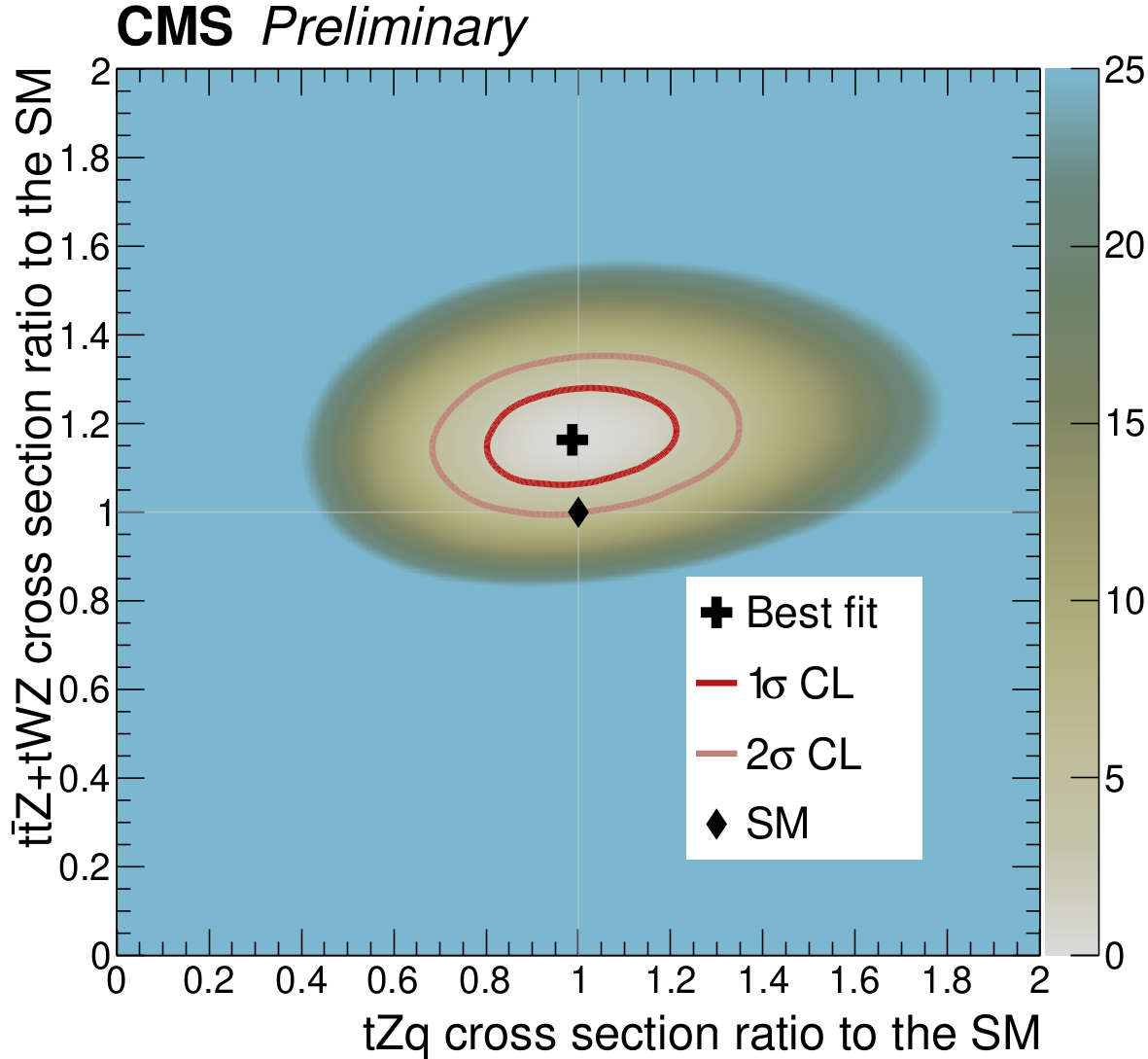}
	\caption{The likelihood function as a function of the $t\bar{t}$Z + $t$WZ cross section ratio to SM and the tZq cross section ratio to SM.}
	\label{fig:inclusive_scan}
\end{figure}

The fit converges at cross section ratios to the SM of 1.17~$\pm$~0.07 for the $t\bar{t}$Z~+~$t$WZ  processes and 0.99~$\pm$~0.13 for the tZq~process. 
The inclusive cross sections of the signal processes are defined in the phase space including resonant and nonresonant production of opposite-sign and same-flavour lepton pairs with an invariant mass 70~<~$m_{\ell^+\ell^-}$~<~110 GeV. 
The predicted cross sections were evaluated in previous CMS measurements from the signal generator {\MADGRAPH}\_a\MCATNLO~v2.6.5. 
For the $t\bar{t}$Z~+~$t$WZ processes, they were calculated to be 840~$\pm$~100~pb and $136^{+9}_{-8}$~fb, respectively~\cite{ttZ_2020, Hayrapetyan_2024}. 
For the tZq process, the expected value was evaluated to be 94.2 $\pm$ 3.1 fb \cite{tZ_2022} in the phase space where Z boson decays into a lepton pair. 
As the calculation also includes nonresonant lepton-pair production with an invariant mass greater than 30~GeV, a transfer factor is evaluated from the simulated samples to get the tZq cross section in the correct phase space. 
The branching ratio is taken into account as well.
The 2-dimensional scan of the likelihood function over the cross section ratios is shown on the Fig.~\ref{fig:inclusive_scan}. 

The final measured inclusive cross sections are: 
\begin{align}
\begin{split}
	\sigma \left( t\bar{t}\text{Z} + t\text{WZ} \right) &= 1.14 \pm 0.05~\text{(stat)} \pm 0.04 ~\text{(syst) pb}, \\
	\sigma \left( t{Zq} \right) &= 0.81 \pm 0.07~\text{(stat)} \pm 0.06~\text{(syst) pb}.
	\label{eq:inclusive}
\end{split}
\end{align}

The differential cross sections are measured separately as five functions of the observables: 
transverse momentum of Z boson $p_T$(Z); 
transverse momentum of lepton originating from W boson decay $p_T$($\ell_W$); 
azimuthal angle between two leptons originating from Z boson decay $\Delta\phi(\ell^+\ell^-)$; 
spatial separation between Z boson and the lepton originating from W boson decay $\Delta R (Z, \ell_W)$;
cosine of the polar angle between Z boson and negatively charged lepton originating from its decay, boosted into the Z boson rest frame $\cos(\theta_Z^{\ast})$.

The measurement is performed at the parton level: objects are defined based on event generator level particles after initial and final state radiation before hadronization. 
Binning for the different observables is chosen by computing a response matrix for both signal samples.
In each bin of the measurement, templates that describe the expected distributions of signals and
backgrounds are created. 
The signal templates are split into subsamples corresponding to the generator-level bins of the observable to unfold, and the events in the signal output nodes are further split into the different categories corresponding to the detector-level bins. 
In the fit, for each bin, two cross section parameters of interest are used to determine the likelihood ratio in that bin. 

\begin{figure}[!htp]
	\centering
\begin{tabular}{cc}
	\includegraphics[width=0.30\textwidth]{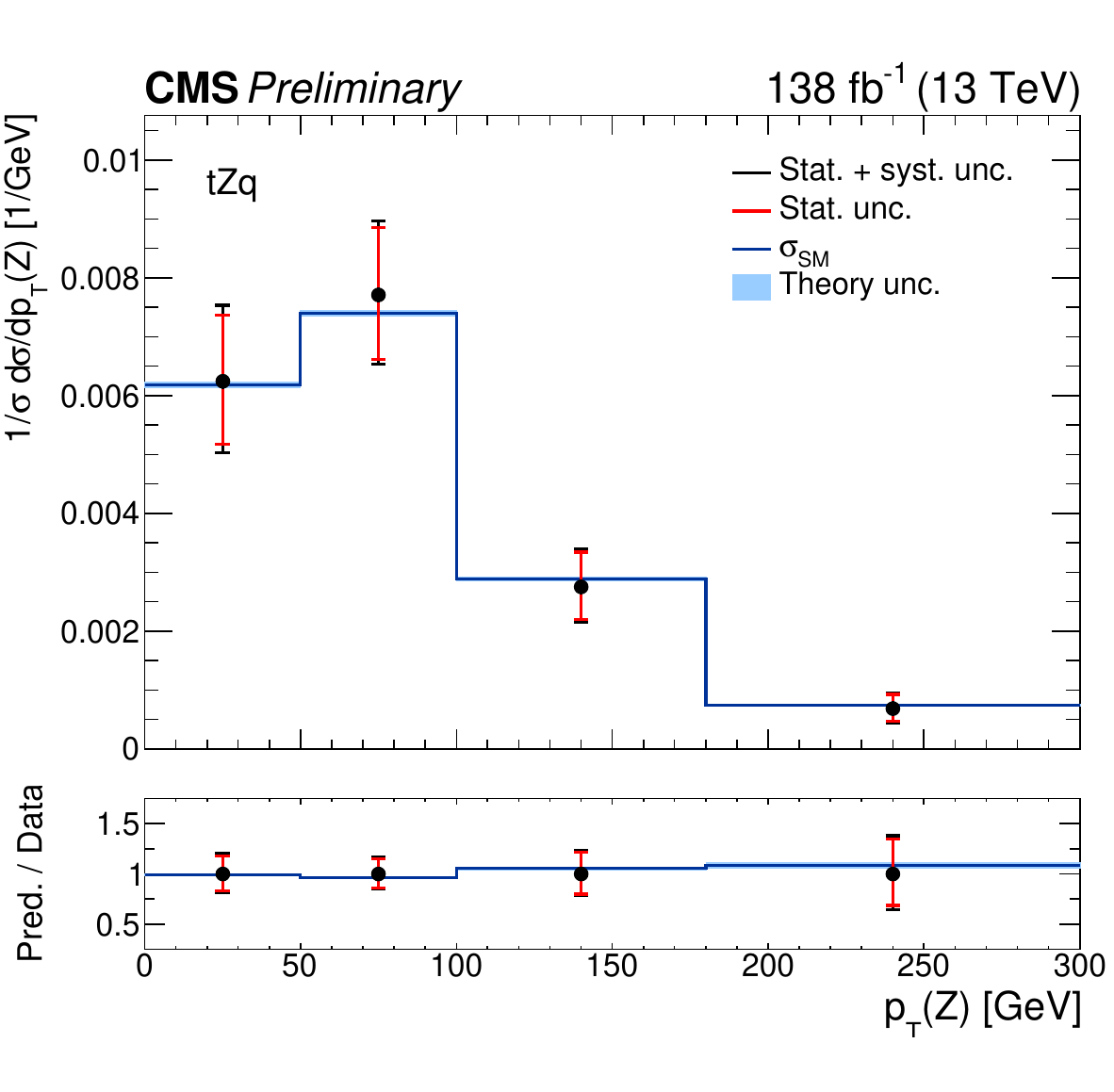} & 
	\includegraphics[width=0.30\textwidth]{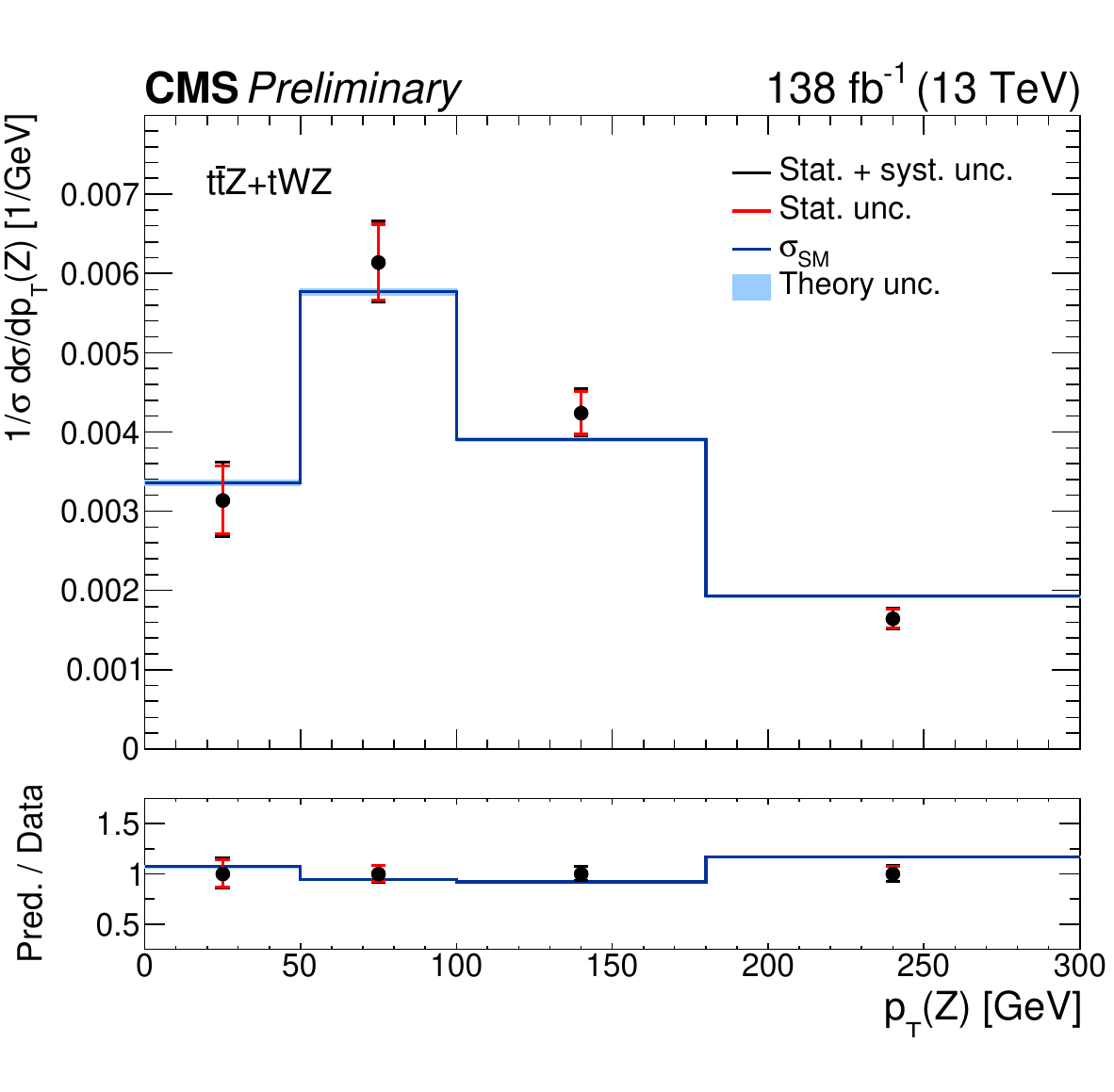} \\
	\includegraphics[width=0.30\textwidth]{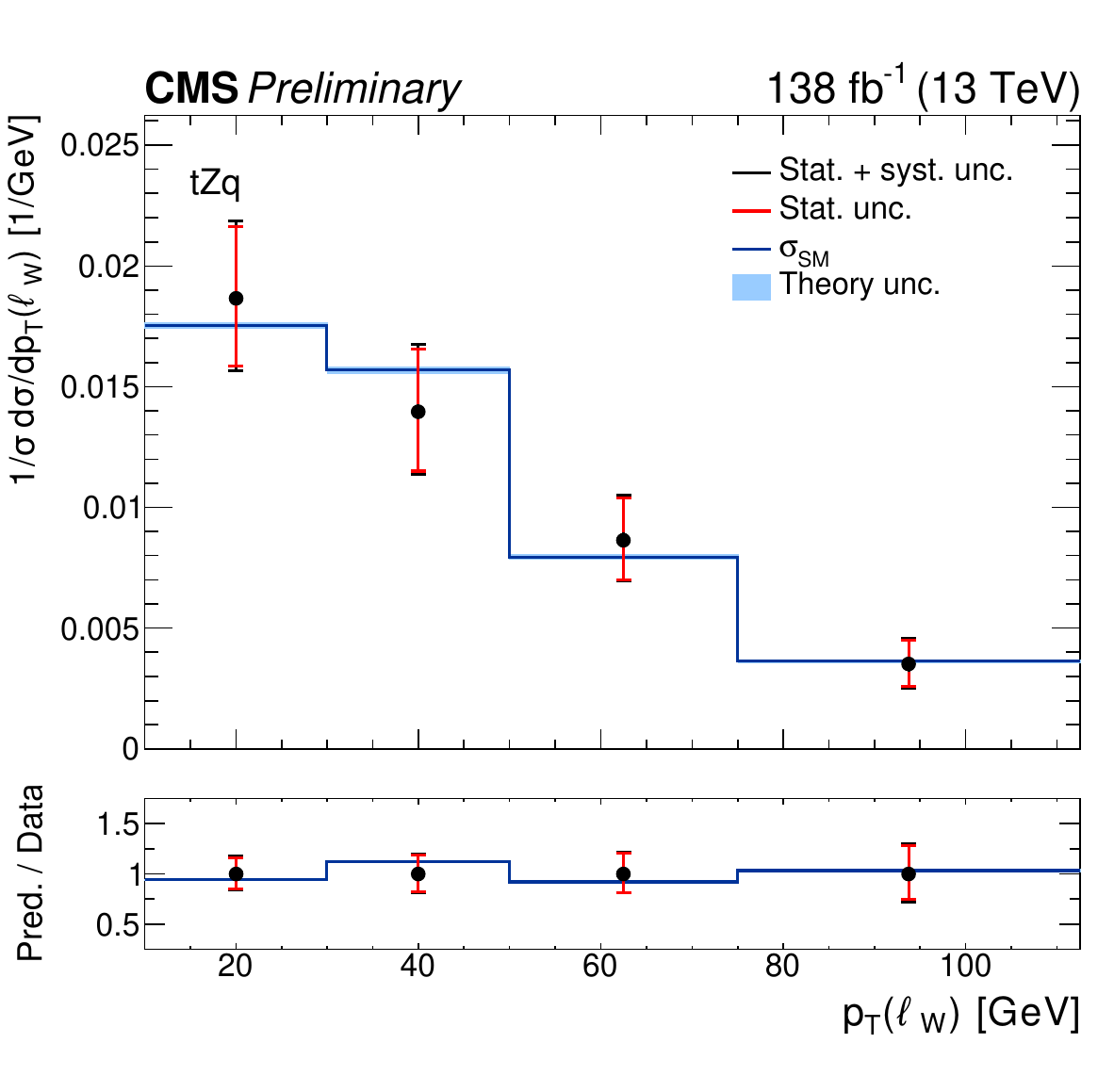} & 
	\includegraphics[width=0.30\textwidth]{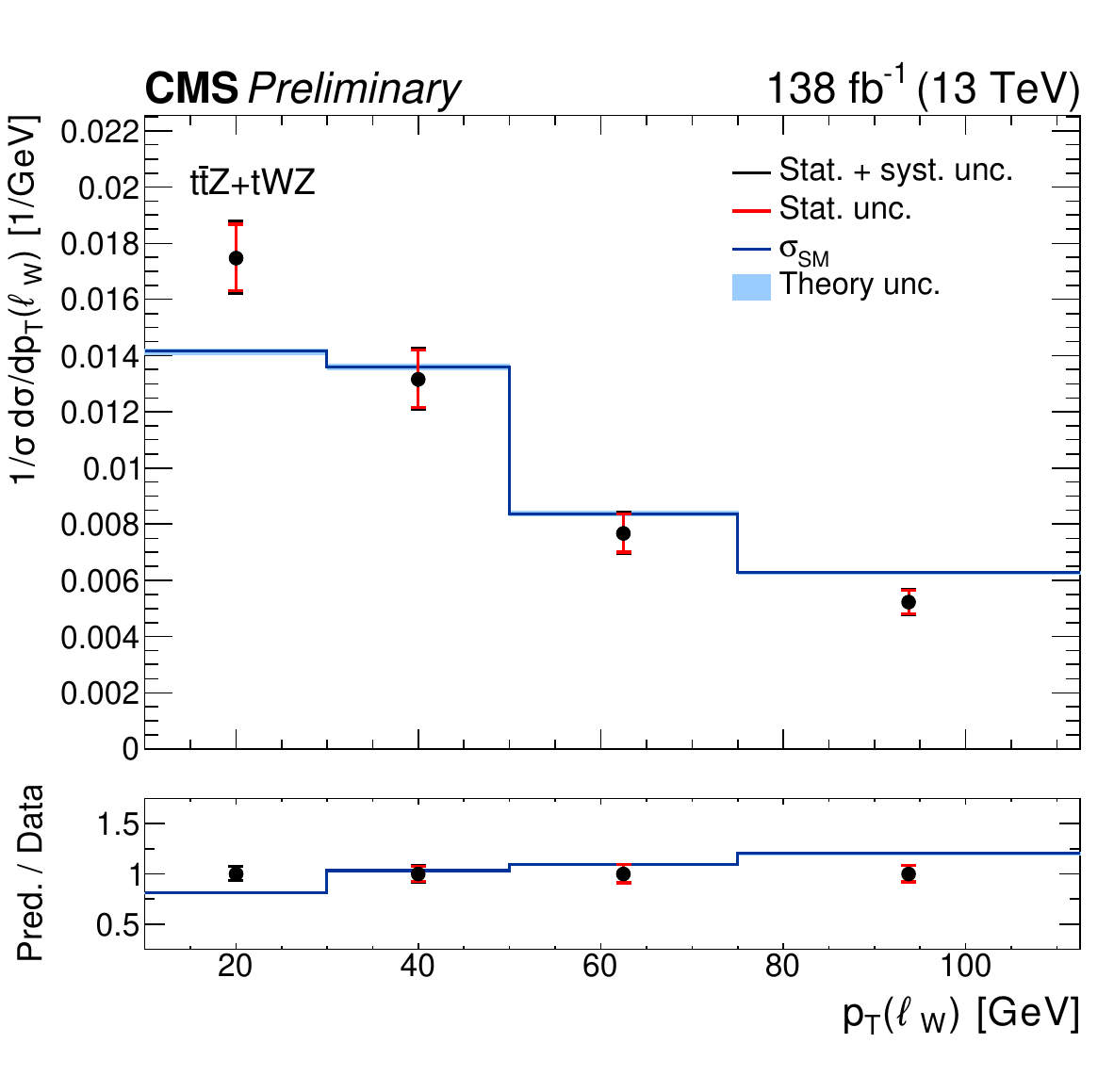} \\
	\includegraphics[width=0.30\textwidth]{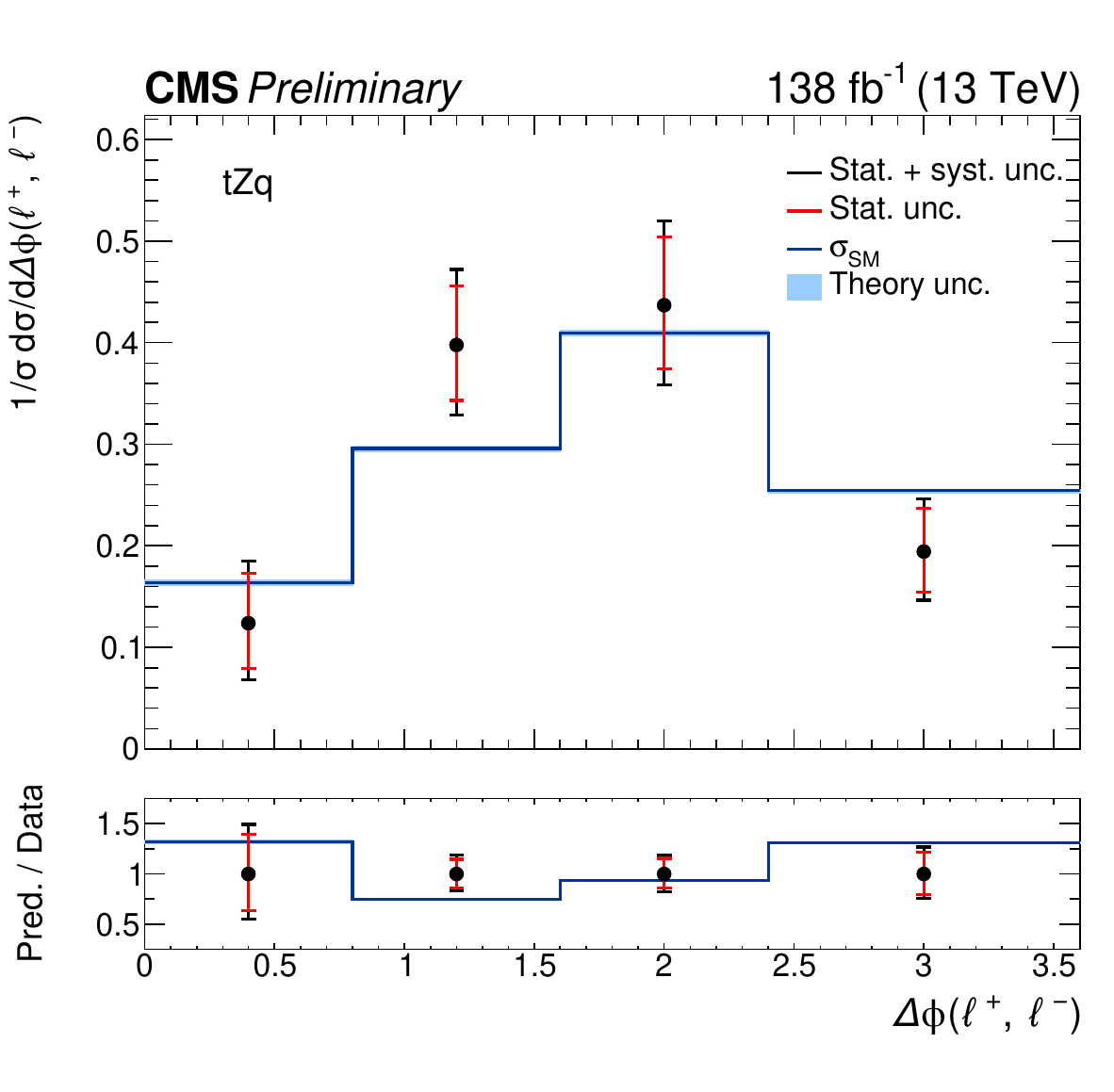} & 
	\includegraphics[width=0.30\textwidth]{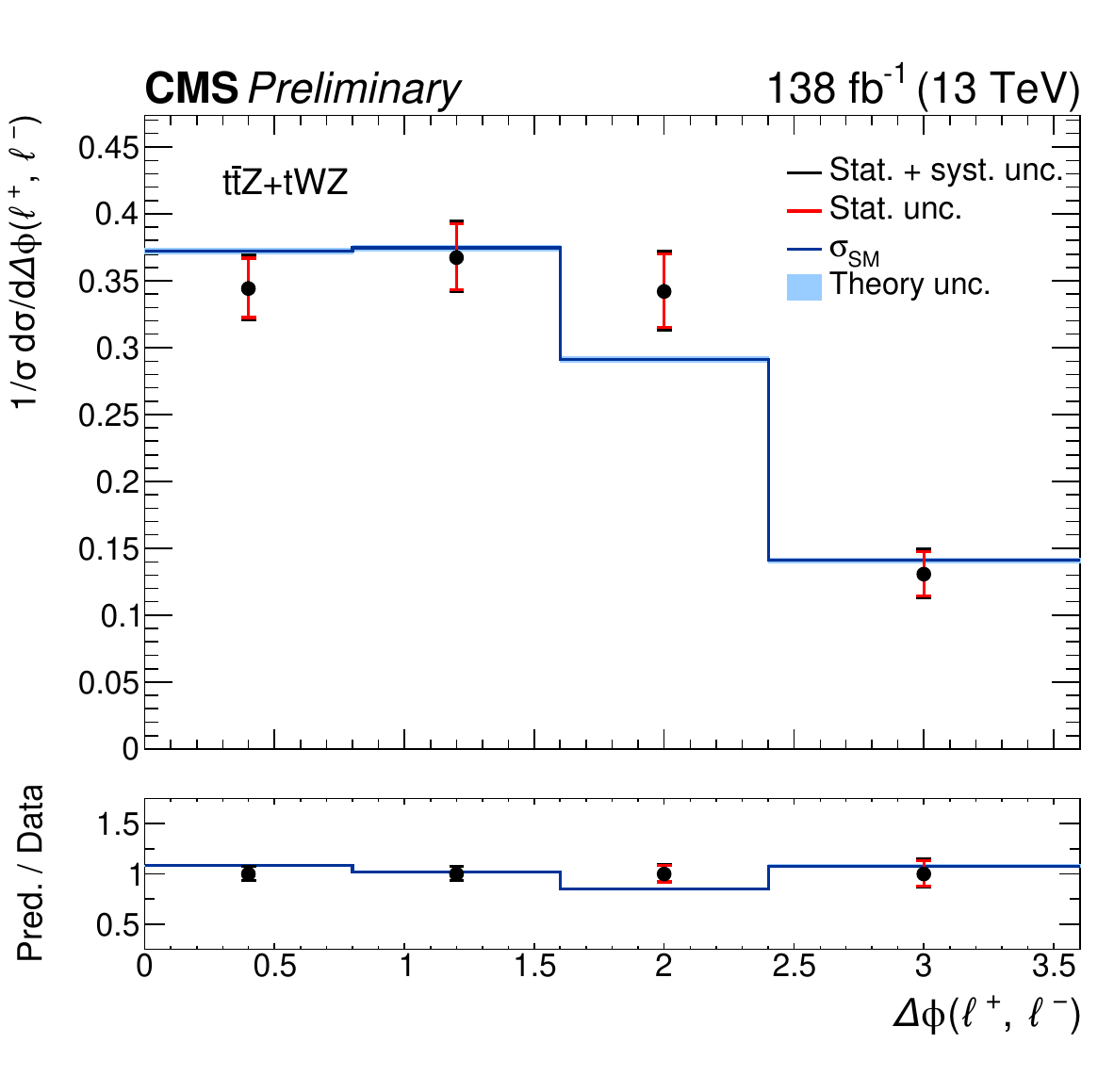} \\
	\includegraphics[width=0.30\textwidth]{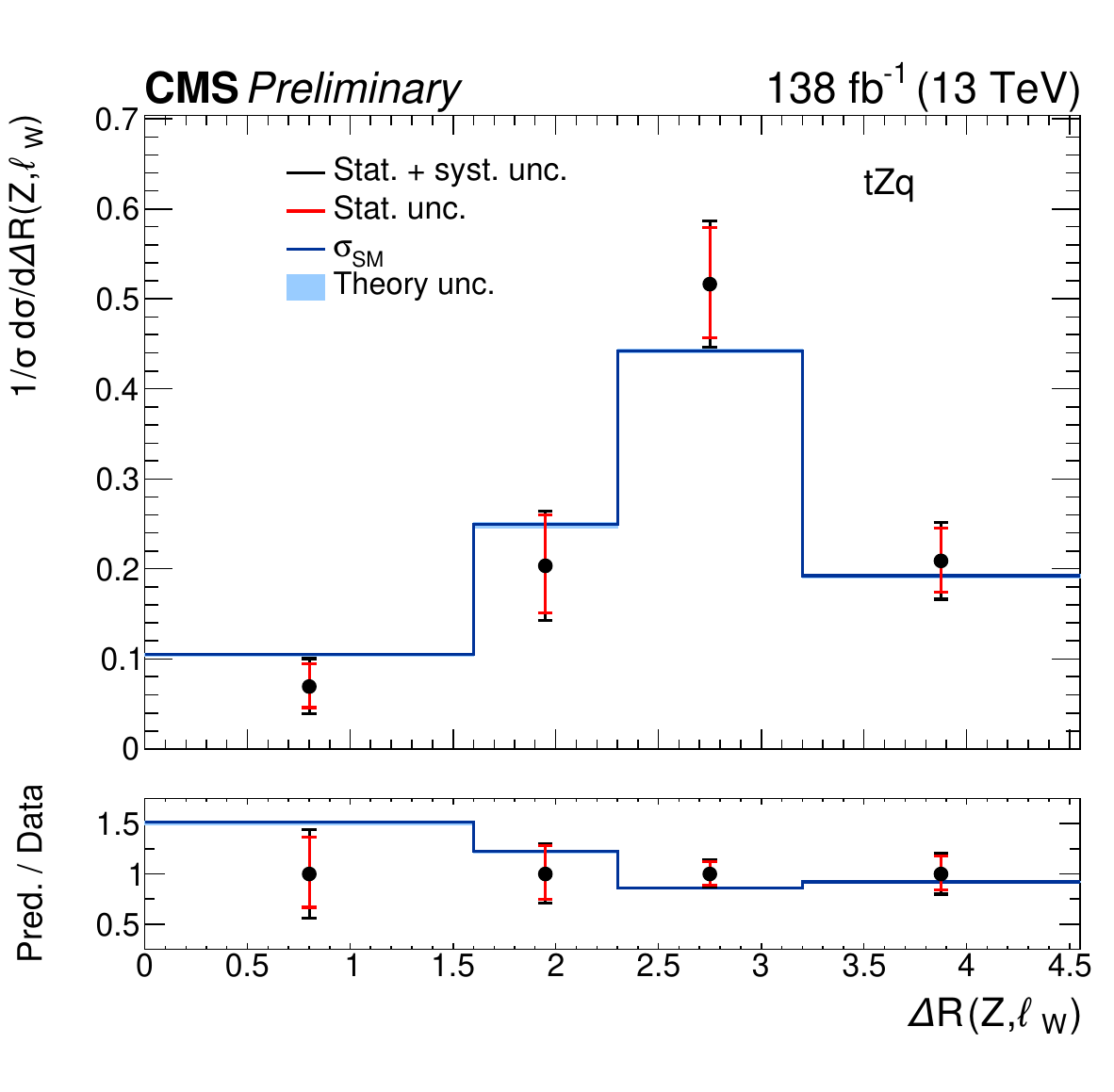} & 
	\includegraphics[width=0.30\textwidth]{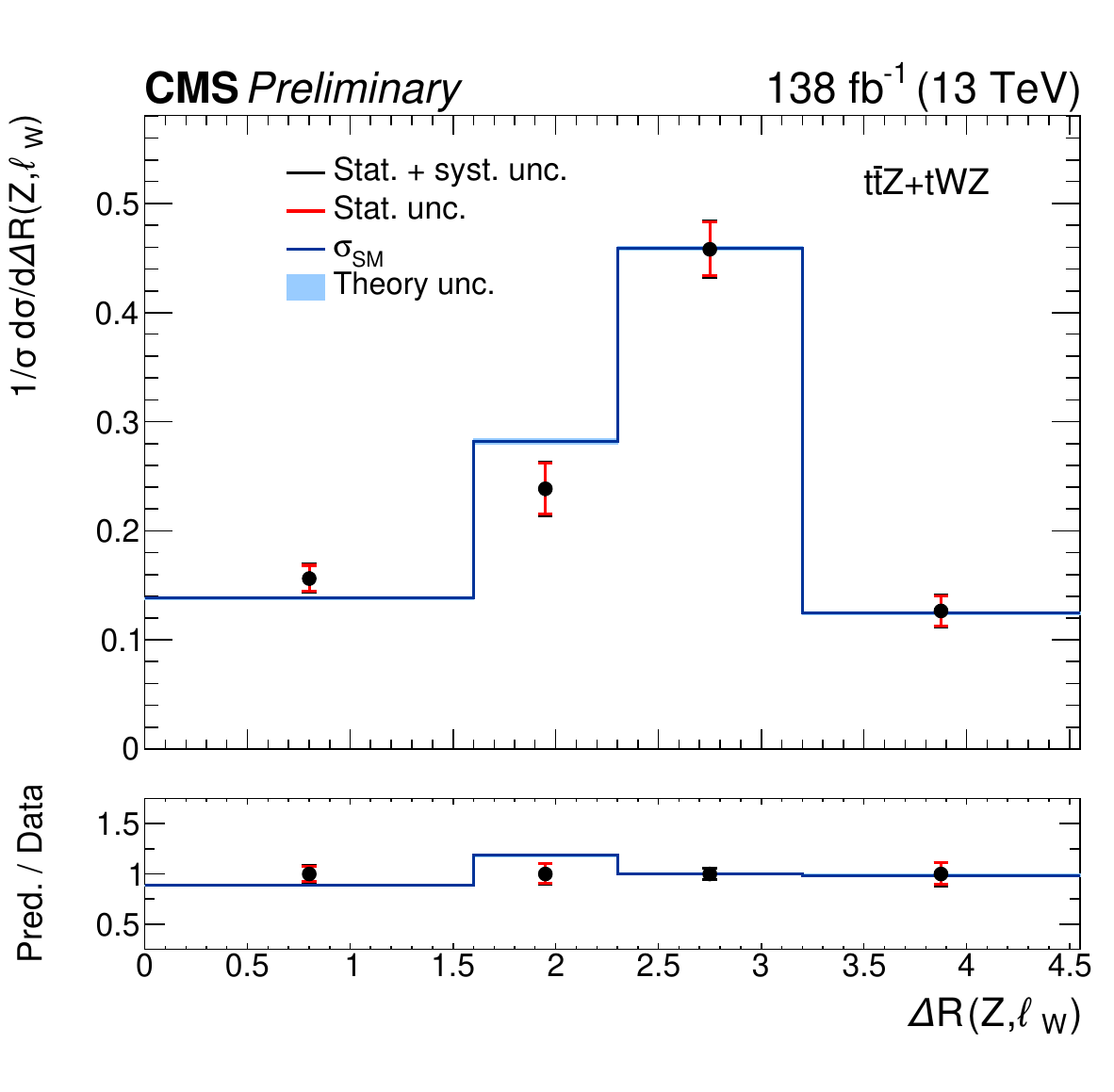} \\
	\includegraphics[width=0.30\textwidth]{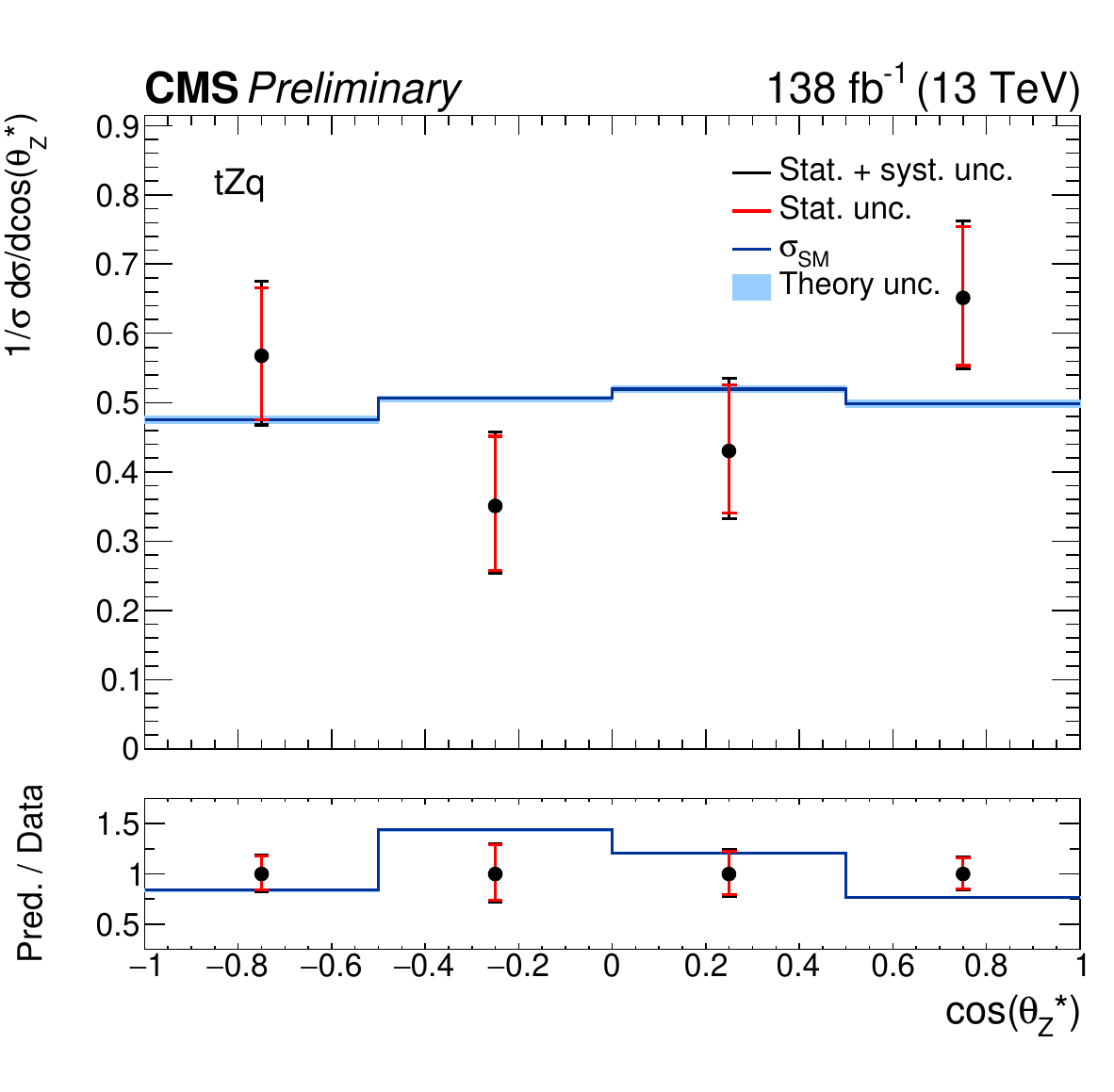} & 
	\includegraphics[width=0.30\textwidth]{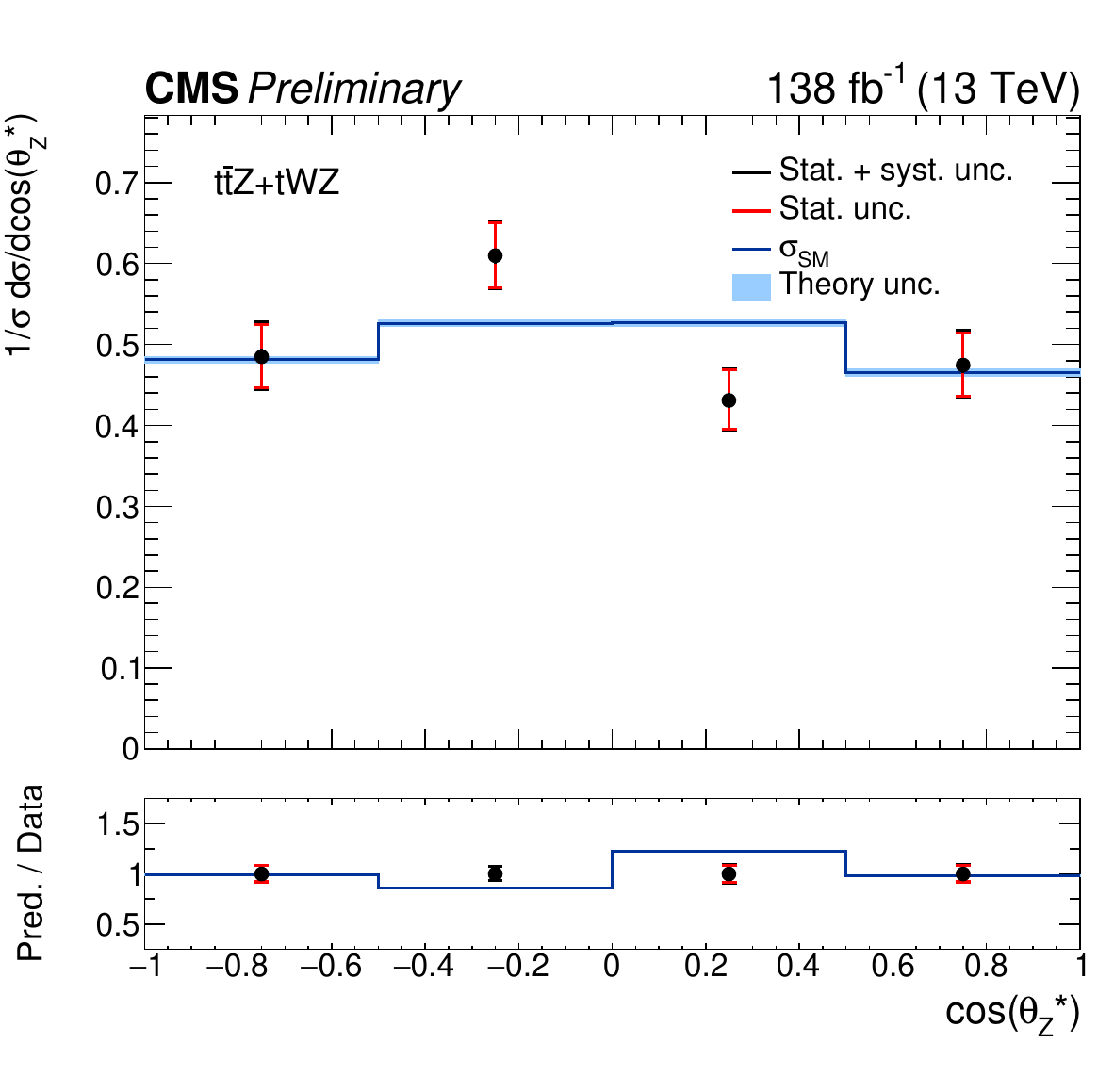} \\
\end{tabular}
\caption{Normalized differential cross sections of the tZq (left column) and the sum of $t\bar{t}Z$ and $t$WZ (right column) as a function of $p_T$(Z) (the first row), $p_T$($\ell_W$) (the second row), $\Delta\phi(\ell^+\ell^-)$ (the third row), $\Delta R (Z, \ell_W)$ (the forth row) and $\cos(\theta_Z^{\ast})$ (the bottom row).}
\label{fig:norm_differential}
\end{figure}

Normalized differential cross sections for tZq and $t\bar{t}$Z~+~$t$WZ processes with the predictions obtained from the MC simulation and the uncertainties are shown on the Fig.~\ref{fig:norm_differential}.
For the sum of $t\bar{t}$Z and $t$WZ, the uncertainties related to the overlap removal are also included. 
The predicted cross sections refer to the same phase space as the inclusive measurement.

The tZq differential cross sections are in good agreement with the theory prediction. 
For the sum of the $t\bar{t}$Z and $t$WZ cross sections, a trend as a function of $p_T$(W) is observed, leading to a discrepancy in the region of low $p_T$ (W). 
The trend is reminiscent of the observation of similar trend in inclusive tt production between the MC simulations at NLO and the data~\cite{differential_CMS2024}.

\section{Conclusion}
The first simultaneous measurement of inclusive and differential cross sections of single and pair production of top quarks in association with Z boson ($t\bar{t}$Z, $t$WZ and tZq) in proton-proton collisions has been presented.
The data recorded by the CMS experiment in 2016-2018 years of data taking at a center-of-mass energy of 13 TeV, corresponding to an integrated luminosity of 138 fb$^{-1}$.
The inclusive cross sections are measured to be $\sigma(t\bar{t}\text{Z} + t\text{WZ}) = 1.14 \pm 0.07$~pb for the sum of $t$WZ and $t\bar{t}$Z processes and $\sigma(tZq) = 0.81 \pm 0.10$ pb for tZq production. 
Results have been obtained for a dilepton invariant mass within 70 and 110 GeV. 
The differential cross section has been measured as functions of five observables.
In general good data-to-simulation agreement has been achieved.
For the sum of the $t\bar{t}$Z and $t$WZ cross sections, a trend as a function of $p_T$(W) is observed, leading to a discrepancy in the region of low $p_T$(W). 
The preprint version of the paper is available at~\cite{colombina_2024}.

\bibliography{topZ_Proceedings_TOP2024.bib}

\end{document}